\documentclass[review]{elsarticle}
\usepackage{amssymb}
\usepackage{lineno,hyperref}
\modulolinenumbers[5]

\journal{Physics of the Dark Universe}
\newcommand{\beq}{\begin{equation}}
\newcommand{\eeq}{\end{equation}}
\newcommand{\bea}{\begin{eqnarray}}
\newcommand{\eea}{\end{eqnarray}}
\newcommand{\eqref}[1]{(\ref{#1})}

\begin{document}
\begin{frontmatter}

\title{Polarisation as a tracer of CMB anomalies: \\ {\sc Planck} results and future forecasts}

\author[mymainaddressLM,mymainaddressAG]{M.~Billi}
\ead{matteo.billi3@unibo.it}
\author[mymainaddressAG,mysecondaryaddressAG]{A.Gruppuso\corref{mycorrespondingauthor}}
\cortext[mycorrespondingauthor]{Corresponding author}
\ead{alessandro.gruppuso@inaf.it}
\author[mymainaddressNM,mymainaddressAG]{N.~Mandolesi}
\ead{mandolesi@iasfbo.inaf.it}
\author[mymainaddressLM,mymainaddressAG,mysecondaryaddressAG]{L.~Moscardini}
\ead{lauro.moscardini@unibo.it}

\author[mymainaddressNM,mymainaddressML]{P.~Natoli}
\ead{paolo.natoli@unife.it}
\address[mymainaddressLM]{Dipartimento di Fisica e Astronomia, Alma Mater Studiorum Universit\`a di Bologna, Via Gobetti 93/2, I-40129 Bologna, Italy}
\address[mymainaddressAG]{INAF-OAS Bologna,
Osservatorio di Astrofisica e Scienza dello Spazio di Bologna,
Istituto Nazionale di Astrofisica,
via Gobetti 101, I-40129 Bologna, Italy}
\address[mysecondaryaddressAG]{INFN, Sezione di Bologna,
Via Irnerio 46, I-40126 Bologna, Italy}
\address[mymainaddressNM]{Dipartimento di Fisica e Scienze della Terra, Universit\`a degli Studi di Ferrara, Via Saragat 1, I-44100 Ferrara, Italy}
\address[mymainaddressML]{INFN, Sezione di Ferrara, Via Saragat 1, I-44100 Ferrara, Italy}

%
\begin{abstract}
\noindent
The lack of power anomaly is an intriguing feature at the largest angular scales of the CMB anisotropy temperature pattern, whose statistical significance is not strong enough to claim any new physics beyond the standard cosmological model.
We revisit the former statement by also considering polarisation data. We propose a new one-dimensional estimator which takes jointly into account the information contained in the TT, TE and EE CMB spectra. By employing this estimator on {\sc Planck} 2015 low-$\ell$ data, we find that a random $\Lambda$CDM realisation is statistically accepted at the level of $3.68 \%$. 
Even though {\sc Planck} polarisation contributes a mere $4 \%$ to the total information budget, its use pushes  the lower-tail-probability down from the $7.22 \%$ obtained with only temperature data. 
Forecasts of future CMB polarised measurements, as e.g. the LiteBIRD satellite, can increase the polarisation contribution up to $6$ times with respect to {\sc Planck} at low-$\ell$.
We argue that the large-scale E-mode polarisation may play an important role in analysing CMB temperature anomalies with future mission.
\end{abstract}
%

\end{frontmatter}

%
%
%
%
\section{Introduction}
\label{intro}

CMB observations show anomalies at large angular scale of the temperature map, see e.g. \cite{Schwarz:2015cma}. The statistical level of these signatures is around 2-3$\sigma$ from what expected in the concordance $\Lambda$CDM model. Not all of these anomalies are independent and a certain degree of correlation exists \cite{Muir:2018hjv}. Here we focus on the lack of power anomaly: the temperature CMB anisotropy pattern exhibits less power with respect to what foreseen by $\Lambda$CDM. This effect has been studied with the variance estimator in WMAP data \cite{Monteserin:2007fv,Cruz:2010ud,Gruppuso:2013xba} and in Planck 2013 \cite{Ade:2013nlj} and Planck 2015 \cite{Ade:2015hxq} data, measuring a lower-tail-probability of the order of few per cent. Such a percentage can become even smaller, below $1 \%$, once only regions at high Galactic latitude are taken into account \cite{Gruppuso:2013xba}.

WMAP and Planck agree well on this feature, so it is very hard, albeit not impossible, to attribute this anomaly to systematic effects of instrumental origin. 
Moreover it is also difficult to believe that a lack of power could be generated by residuals of astrophysical emission, since the latter is not expected to be correlated\footnote{In particular they should be anti-correlated to produce a decrease of the total power.} with the CMB and therefore an astrophysical residual should increase the total power rather than decreasing it.
Hence, it appears natural to accept this as a real feature present in the CMB pattern. 

An early fast-roll phase of the inflaton could naturally explain such a lack of power, see e.g. \cite{Contaldi:2003zv,Cline:2003ve,Destri:2009hn,Cicoli:2013oba,augusto_cmbdepression}: this anomaly might then witness a new cosmological phase before the standard inflationary era (see e.g. \cite{Gruppuso:2015zia,Gruppuso:2015xqa,Gruppuso:2017nap}). 

However, with only the observations based on the temperature map, this anomaly is not statistically significant enough to be used to claim new physics beyond the standard cosmological model. Therefore, it is legitimate to conservatively interpret it as a statistical fluke of the $\Lambda$CDM concordance model.  

The main point of this paper is to argue that future CMB polarisation data at low-$\ell$ might increase the significance of this anomaly. 
In other words, considering the counterpart in polarisation of the lack of power currently observed in temperature might be key to confirm it as a simple statistical fluke or to raise it up at the level of manifestation of new physics\footnote{Of course this argument might be used for any CMB anomaly.}. 

In this paper we propose a new one-dimensional estimator which combines information from the CMB TT, EE and TE angular power spectra at the largest angular scales, i.e. $2 \le \ell \le 30$, with $\ell$ being 
the multipole moment. 
Considering {\sc Planck} data in the whole harmonic range mentioned above, noise dominated polarisation provides an information content at the level of $4 \%$ to this estimator which, even though small, has a non-negligible impact on the analysis, the lower-tail-probability shifting downward from $7.22 \%$ (obtained considering only temperature data) to $3.68 \%$ C.L. (obtained considering jointly temperature and polarisation data).
We show that for future CMB observations, polarisation at the largest angular scales can weight as much as $\sim 23 \%$ of the total information entering our estimator.

We argue that the inclusion of large-scale E-mode polarisation could crucially help in changing the interpretation from a simple statistical fluke into the detection of a new physical phenomenon. 
Therefore, future CMB large-scale polarised observations, which are typically aimed at primordial B-modes, might provide signals of new physics also through the other polarised CMB mode, i.e. the E-mode. 

The paper is organised as follows: in Section \ref{estimator} we introduce the algebra needed to build the new estimator which condensates all the TT, EE and TE information into a 1-D object; in Section \ref{optmised} we elaborate on an optimised (i.e. minimum variance) version of the proposed estimator; Section \ref{dataandsims} is devoted to the description of the dataset used and of the simulations employed; in Section \ref{results} we present the results on {\sc Planck} data and provide estimates of the improvement expected with future CMB polarised observations, as the LiteBIRD satellite \cite{Suzuki:2018cuy}; 
conclusions are drawn in Section \ref{conclusions}.

\section{A new one-dimensional joint estimator: the dimensionless normalised mean power}
\label{estimator}

The idea of this joint estimator starts from the usual equations employed to simulate temperature and E-mode CMB maps, see e.g.~\cite{Copi:2013zja}:
\begin{eqnarray} 
a_{\ell m}^{T} & = & \sqrt{C_{\ell}^{TT, th}} \xi_{\ell m}^{1}  \, , \label{almT} \\ 
a_{\ell m}^{E} & = & \frac{C_{\ell}^{TE,th}}{\sqrt{C_{\ell}^{TT,th}}} \xi_{\ell m}^{1}  +  \sqrt{C_{\ell}^{EE,th} - \frac{(C_{\ell}^{TE,th})^{2}}{C_{\ell}^{TT,th}}} \xi_{\ell m}^{2} \, , \label{almE}
\end{eqnarray}
where $a_{\ell m}^{T,E}$ are the coefficients of the Spherical Harmonics (with $\ell, m$ being integers numbers so that $\ell \in \{0,1,2,3...\}$ and $-\ell \le m \le \ell $), $C_{\ell}^{TT, th}$, $C_{\ell}^{EE, th}$ and $C_{\ell}^{TE, th}$ are the theoretical angular power spectra (APS) for $TT$, $EE$ and $TE$ and with $\xi_{\ell m}^{1,2}$ being Gaussian random variables,  uncorrelated, with zero mean and unit variance:
\begin{eqnarray}
\langle \xi_{\ell m}^{1} \rangle & = & 0 \, , \label{meanxi1} \\
\langle \xi_{\ell m}^{2} \rangle & = & 0 \, , \label{meanxi2} \\
\langle \xi_{\ell m}^{1} \xi_{\ell' m'}^{2} \rangle & = & 0 \, , \label{meanxi1xi2}\\
 \langle \xi_{\ell m}^{1} \xi_{\ell' m'}^{1} \rangle & = & \langle \xi_{\ell m}^{2} \xi_{\ell' m'}^{2} \rangle =  \delta_{\ell \ell'} \,  \delta_{m m'} \, . \label{variancexi1xi2}
\end{eqnarray}
From equations (\ref{almT}),(\ref{almE}) one can compute the corresponding APS, defined as
\begin{eqnarray} 
C_{\ell}^{TT,sim} & = & \frac{1}{2 \ell +1} \sum_{m=-\ell}^{\ell} a^{T}_{\ell m} (a^{T}_{\ell m} )^{\star} \, , \\ 
C_{\ell}^{TE,sim} & = & \frac{1}{2 \ell +1} \sum_{m=-\ell}^{\ell} a^{T}_{\ell m} (a^{E}_{\ell m} )^{\star} \, , \\ 
C_{\ell}^{EE,sim} & = & \frac{1}{2 \ell +1} \sum_{m=-\ell}^{\ell} a^{E}_{\ell m} (a^{E}_{\ell m} )^{\star} \, ,
\end{eqnarray}
where the label $sim$ stands for ``simulated'', i.e. realised randomly from the theoretical spectra $C_{\ell}^{TT, th}$, $C_{\ell}^{EE, th}$ and $C_{\ell}^{TE, th}$, finding the following expressions,
\begin{eqnarray}
C_{\ell}^{TT,sim}  & = & C_{\ell}^{TT,th} \frac{ | \vec{\xi}^{(1)}_{\ell} |^{2}}{2 \ell+1} \, , \label{CellTT} \\
C_{\ell}^{EE,sim} & = & \frac{(C_{\ell}^{TE,th})^{2}}{C_{\ell}^{TT,th}} \left[\frac{| \vec{\xi}^{(1)}_{\ell} |^{2}}{2\ell+1} - \frac{|\vec{\xi}^{(2)}_{\ell}|^2}{2\ell+1}\right] + C_{\ell}^{EE,th} \frac{| \vec{\xi}^{(2)}_{\ell} |^{2}}{2 \ell+1} \nonumber\\
                        && + 2 a_{\ell}  \frac{{C_{\ell} ^{TE,th}}}{{C_{\ell}^{TT,th}}} \frac{ \vec{\xi}^{(1)}_{\ell} \cdot \vec{\xi}^{(2)}_{\ell} }{2 \ell+1} \, , \label{CellEE} \\
C_{\ell}^{TE,sim} & = & C_{\ell}^{TE,th}\frac{| \vec{\xi}^{(1)}_{\ell} |^{2}}{2\ell+1} +a_{\ell} \frac{\vec{\xi}^{(1)}_{\ell} \cdot \vec{\xi}^{(2)}_{\ell}}{2\ell+1}  \, , \label{CellTE}
\end{eqnarray}
where $\vec{\xi}^{(1/2)}_{\ell}$ are vectors with $2 \ell +1$ components, i.e.
\begin{equation}
\vec{\xi}^{(1/2)}_{\ell} = \left( \xi_{-\ell}^{(1/2)}, \xi_{-\ell+1}^{(1/2)}, ... , \xi_{0}^{(1/2)}, ... \xi_{\ell-1}^{(1/2)}, \xi_{\ell}^{(1/2)} \right) \, ,
\end{equation}
and $a_{\ell}$ is defined as
\begin{equation}
a_{\ell}  \equiv  \sqrt{C_{\ell}^{EE,th} C_{\ell}^{TT,th} -  (C_{\ell}^{TE,th})^{2}} \, . \label{a}
\end{equation}

It is easy to check that taking the ensemble average of equations (\ref{CellTT}),(\ref{CellEE}) and (\ref{CellTE}) yields to
\begin{eqnarray} 
\langle C_{\ell}^{TT,sim} \rangle & = & C_{\ell}^{TT,th} \, , \\ 
\langle C_{\ell}^{EE,sim} \rangle  & = & C_{\ell}^{EE,th} \, , \\ 
\langle C_{\ell}^{TE,sim} \rangle & = & C_{\ell}^{TE,th}  \, ,
\end{eqnarray}
since for each $\ell$, as a consequence of equations (\ref{meanxi1xi2}),(\ref{variancexi1xi2}),
\begin{eqnarray} 
\langle \frac{ | \vec{\xi}^{(1)}_{\ell} |^{2}}{2 \ell+1} \rangle & = & 1 \, , \label{meanxi1} \\ 
\langle \frac{ | \vec{\xi}^{(2)}_{\ell} |^{2}}{2 \ell+1} \rangle & = & 1 \, , \label{meanxi2} \\ 
\langle \vec{\xi}^{(1)}_{\ell} \cdot \vec{\xi}^{(2)}_{\ell} \rangle & = & 0  \, . \label{meanxi1dotxi2} 
\end{eqnarray}

Equations (\ref{CellTT}),(\ref{CellEE}) and (\ref{CellTE}) can be inverted, giving the following set of equations
\begin{eqnarray}
 \frac{ | \vec{\xi}^{(1)}_{\ell} |^{2}}{2 \ell+1} & = & \frac{C_{\ell}^{TT} }{C_{\ell}^{TT,th}} \, , \\
 \frac{ | \vec{\xi}^{(2)}_{\ell} |^{2}}{2 \ell+1} & = & \frac{C_{\ell}^{EE}}{a_{\ell}^{2}} C_{\ell}^{TT,th} - \frac{C_{\ell}^{TT,th}}{a_{\ell}^{2}} \left(\frac{C_{\ell}^{TE,th} }{C_{\ell}^{TT,th}}\right)^{2} C_{\ell}^{TT} \nonumber \\                                                                      
                                                                         && - 2 \frac{C_{\ell}^{TE,th}}{a_{\ell}^{2}}\left[C_{\ell}^{TE} - \frac{C_{\ell}^{TE,th} }{C_{\ell}^{TT,th}} C_{\ell}^{TT}\right] \, , \\
\frac{ \vec{\xi}^{(1)}_{\ell} \cdot \vec{\xi}^{(2)}_{\ell}}{2 \ell +1}     & = & \frac{1}{a_{\ell}} \left[C_{\ell}^{TE} - \frac{C_{\ell}^{TE,th} }{C_{\ell}^{TT,th}} C_{\ell}^{TT}\right] \, ,
\end{eqnarray}
where we have dropped out the label ``$sim$'' for sake of simplicity. 
Now, we can interpret $C_{\ell}^{TT}$, $C_{\ell}^{EE}$ and $C_{\ell}^{TE}$ as the CMB APS recovered by a CMB experiment under realistic circumstances, i.e. including noise residuals, incomplete sky fraction and finite angular resolution\footnote{In principle one can also include residuals of systematic effects.}.
Once the model is chosen, i.e. once the spectra $C_{\ell}^{TT, th}$, $C_{\ell}^{EE, th}$ and $C_{\ell}^{TE, th}$ are fixed, for example to $\Lambda$CDM, one can compute the following objects
\begin{eqnarray}
x^{(1)}_{\ell} & \equiv & \frac{ | \vec{\xi}^{(1)}_{\ell} |^{2}}{2 \ell+1}  \, , \\
x^{(2)}_{\ell} & \equiv & \frac{ | \vec{\xi}^{(2)}_{\ell} |^{2}}{2 \ell+1}  \, , \\
x^{(3)}_{\ell} & \equiv & \frac{ \vec{\xi}^{(1)}_{\ell} \cdot \vec{\xi}^{(2)}_{\ell}}{2 \ell +1}  \, ,
\end{eqnarray}
for the observations and/or for the corresponding realistic simulations.

In the following we will call the variables $x^{(1)}_{\ell}$, $x^{(2)}_{\ell}$ and $x^{(3)}_{\ell}$ as APS of the normal random variables or normalised APS (henceforth NAPS). 
The advantage of using NAPS, instead of the standard APS, is that they are dimensionless and similar amplitude numbers and can be easily combined to define a 1-D estimator in harmonic space, which depends on temperature, E-mode polarisation and their cross-correlation. A natural definition of this 1-D estimator, called $P$, is the following
\begin {equation}
P = \frac{1}{(\ell_{max}-1)} \sum^{\ell_{max}}_{\ell=2} \left( x_{\ell}^{(1)} + x_{\ell}^{(2)} \right). \label{definitionofP}
\end{equation}
The estimator $P$ could be interpreted as a dimensionless normalised mean power, which jointly combines the temperature and polarisation data. The expectation value of $P$ is
\begin{equation}
\langle P \rangle = 2 \, , 
\label{expectedvalueP}
\end{equation}
regardless of the value of $\ell_{max}$. Note that a definition of the following type
\begin{equation}
S = \frac{1}{(\ell_{max} -1)}\sum_{\ell=2}^{\ell_{max}}  ( x^{(1)}_{\ell} + x^{(2)}_{\ell} + x^{(3)}_{\ell} ) \, ,
\end{equation}
is expected to have less signal-to-noise ratio with respect to $P$ because while $S$ and $P$ have the same expectation value, the intrinsic variance of $P$ is in general smaller than the one of $S$.

We will see in the following that Eq.~(\ref{definitionofP}) is noise-limited for Planck data due to $x^{(2)}_{\ell}$ (its polarisation part) and in practice can be employed only up to $\ell_{max}=6$. 
An optimised version of this estimator, given in Section \ref{optmised}, does not suffer from this issue and can be employed up to the maximum multipole considered in this analysis, i.e. $\ell_{max}=30$.

\section{Optimised estimator}
\label{optmised}

In equation (\ref{definitionofP}) the NAPS $x^{(1)}_{\ell}$ and $x^{(2)}_{\ell}$ are combined with equal weights. 
However the signal-to-noise ratios of the two NAPS are different even in the cosmic variance limit case: therefore one might wonder which are the best weights that we can use in the definition 
of the joint estimator in order to make it optimal, i.e. with minimum variance. 
It is possible to show, see \ref{optmisedweights}, that the optimised estimator $\tilde P$, defined as
\begin{equation}
\tilde P \equiv {1 \over (\ell_{max}-1) }\sum_{\ell=2}^{\ell_{max}} ( \alpha_{\ell} x^{(1)}_{\ell} + \beta_{\ell} x^{(2)}_{\ell} ) 
\label{definitionofPtilde}
\end{equation}
has minimum variance when
\begin{eqnarray}
\alpha_{\ell} = 2 {var(x^{(2)}_{\ell})-cov(x^{(1)}_{\ell} ,x^{(2)}_{\ell}) \over var( x^{(1)}_{\ell} ) + var( x^{(2)}_{\ell} )-2cov(x^{(1)}_{\ell} ,x^{(2)}_{\ell}) } \, , \label{alphasol} \\
\beta_{\ell} = 2 {var( x^{(1)}_{\ell} )-cov(x^{(1)}_{\ell} ,x^{(2)}_{\ell}) \over var( x^{(1)}_{\ell} ) + var( x^{(2)}_{\ell} )-2cov(x^{(1)}_{\ell} ,x^{(2)}_{\ell}) } \, , \label{betasol}
\end{eqnarray}
where $var$ and $cov$ stand respectively for the variance and the covariance of the variables which appear in the brackets.
These coefficients, namely $\alpha_{\ell}$ and $\beta_{\ell}$ as defined in Eqs.(\ref{alphasol}-\ref{betasol}), will be actually used to build the $\tilde P$ estimator.
Note that as done for $P$, $\tilde P$, which depends on $\ell_{max}$, has been normalised such that
$ \langle \tilde P \rangle = 2 $ for any value of $\ell_{max}$.
Note also, that $\tilde P$ can be employed up to $\ell_{max}=30$ both for Planck and LiteBIRD-like simulated data: what changes between the two cases is the set of the coefficients $\alpha_{\ell}$ and $\beta_{\ell}$, or, in other words, the relative contribution of the temperature and polarisation data.

\section{Dataset and simulations}
\label{dataandsims}

We use the latest public {\it Planck} satellite CMB temperature data\footnote{http://www.cosmos.esa.int/web/planck/pla.} i.e. the Planck 2015 Commander map with its standard mask ($f_{sky}^T=93.6$) entering the temperature sector of the low-$\ell$ {\sc Planck} likelihood\footnote{At the moment of writing the corresponding {\sc Planck} 2018 likelihood code and corresponding data set is not publicly available.} \cite{Aghanim:2015xee}. In polarisation we consider a noise-weighted combination of WMAP9 and {\sc Planck} data as done in \cite{Lattanzi:2016dzq}. This allows to gain some signal-to-noise ratio and to deal with a larger sky fraction in polarisation ($f_{sky}^P=73.9$). Temperature and polarisation maps are sampled at HEALPix\footnote{http://healpix.sourceforge.net/.} \cite{Gorski:2004by} resolution $N_{side}=16$. For sake of simplicity we will refer to this data set as the {\sc Planck}-WMAP low-$\ell$ data set.

\begin{figure}[ht]
\centering
\begin{tabular}{c}
\includegraphics[width=103mm]{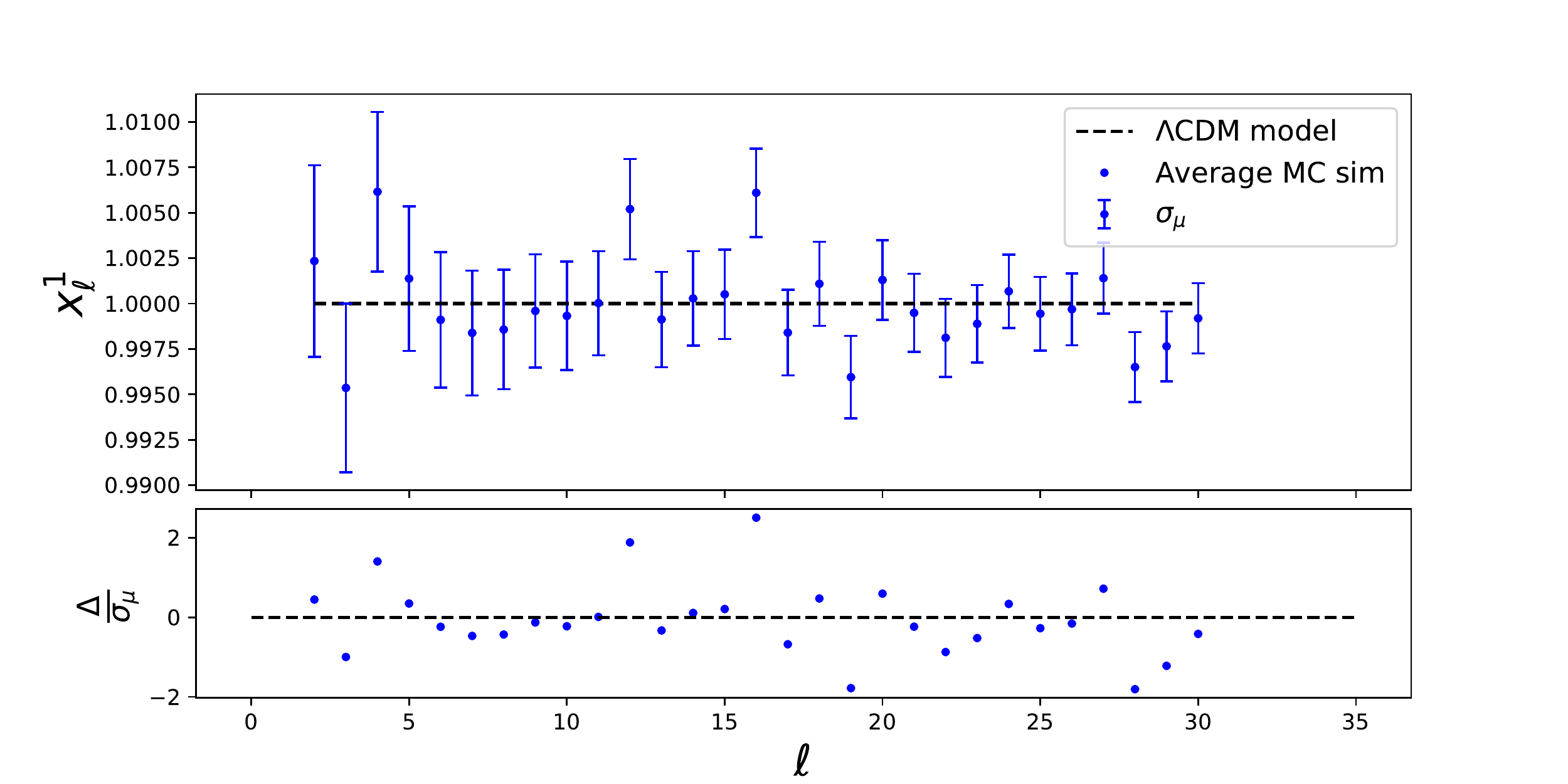}  \\
\includegraphics[width=103mm]{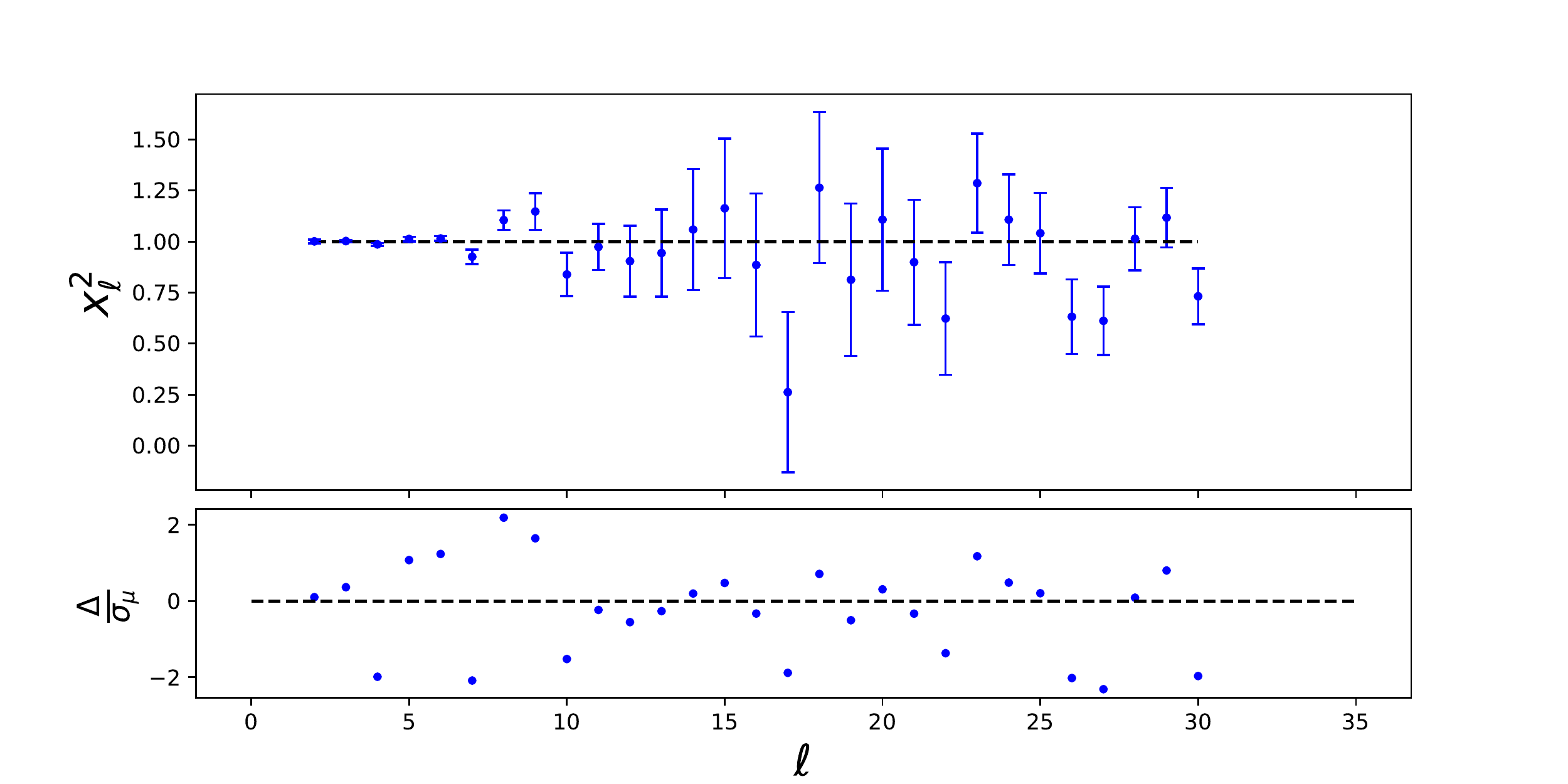} \\
\includegraphics[width=103mm]{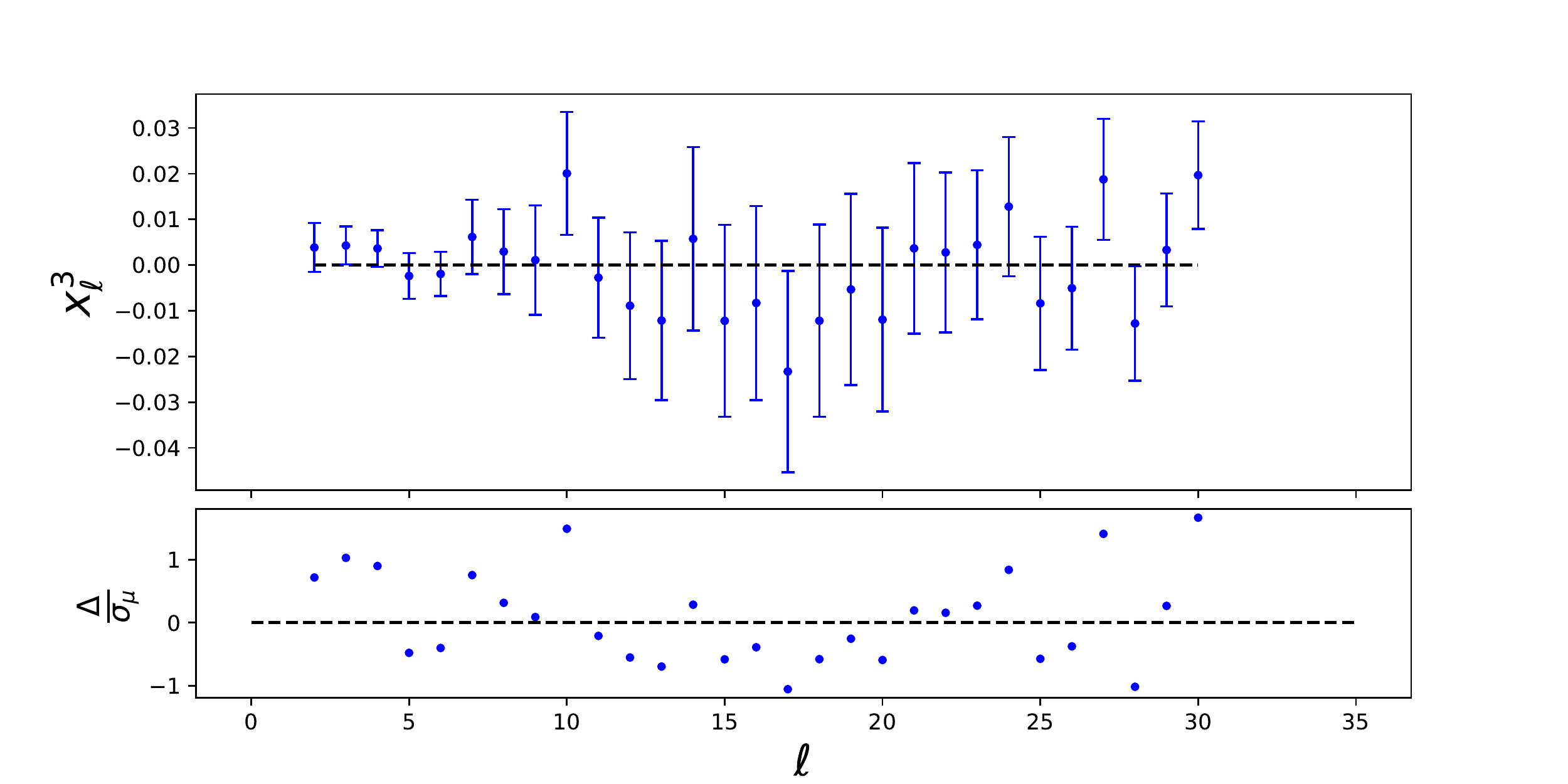}
\end{tabular}
\caption{\small Averages of $x^{(1)}_{\ell}$ (upper panel), $x^{(2)}_{\ell}$ (middle panel) and $x^{(3)}_{\ell}$ (lower panel) as a function of $\ell$ obtained from MonteCarlo simulations corresponding to the {\sc Planck}-WMAP low-$\ell$ data. Error bars represent the uncertainties associated to the averages.
Each panel displays also a lower box where for each $\ell$ it is shown the distance of mean in units of standard deviation of the mean itself. Dashed horizontal lines represent what theoretically expected for 
the averages of $x^{(1)}_{\ell}$, $x^{(2)}_{\ell}$ and $x^{(3)}_{\ell}$, see equations (\ref{meanxi1}),(\ref{meanxi2}) and (\ref{meanxi1dotxi2}).}
\label{fig:one}
\end{figure}

In order to build the estimators $P$ and $\tilde P$, as defined in equations (\ref{definitionofP}) and (\ref{definitionofPtilde}), we estimate the six CMB APS from $10000$ simulated CMB-plus-noise maps. 
The signal is extracted from the {\sc Planck} fiducial $\Lambda$CDM model defined by the following parameters 
\begin{eqnarray}
\Omega_b h^2 = 0.02224 \nonumber \\
\Omega_c h^2 = 0.1187 \nonumber \\
100 \, \theta = 1.04101 \nonumber \\
\tau = 0.065 \nonumber \\
\log[ 10^{10} \, A_s] = 3.060 \nonumber \\
n_s = 0.9673 \nonumber
\end{eqnarray}
being $\Omega_b$ the baryon density, $\Omega_c$ the cold dark matter density, $\theta$ the angle subtended by the sound horizon at recombination, $\tau$ the re-ionization optical depth, $A_s$ the amplitude and $n_s$ the spectral index of primordial scalar perturbations. This set of parameters is obtained confronting data and models through the likelihood function defined as the sum of the three following likelihoods (see  \cite{Aghanim:2015xee,Lattanzi:2016dzq} for further details): 
\begin{itemize} 
\item a pixel based low-$\ell$ likelihood, $2 \le \ell \le 29$, where the Planck 2015 Commander map enters the temperature sector and a noise-weighted combination of WMAP9 and {\sc Planck} data enters the polarisation sector of this likelihood\footnote{Note that the low-$\ell$ data-set used to perform the analysis is exactly the same as the one entering the low-$\ell$ likelihood. This makes the whole investigation self-consistent.};
\item a high-$\ell$ Planck TT likelihood based on APS of Planck data in the range $30 \le \ell \le 2500$;
\item the Planck lensing likelihood, based on the range $40 \le \ell \le 400$ of the four-point correlation function of the temperature anisotropies.
\end{itemize}
The noise of the simulated maps is generated through Cholesky decomposition of the total noise covariance matrix in pixel space, see \ref{choldecomp}. Such estimates are obtained over the observed sky fraction, with the optimal angular power spectrum estimator {\it BolPol} \cite{Gruppuso:2009ab}. 
Montecarlo simulations for the {\sc Planck}-WMAP low-$\ell$ data set are validated in Figure \ref{fig:one}, where the average of the NAPS ($x^{(1)}_{\ell}$, $x^{(2)}_{\ell}$ and $x^{(3)}_{\ell}$) are shown respectively in the upper, middle and lower panels along with their uncertainties of the means ($\sigma_{\mu}$).
Each panel displays also a lower box where for each $\ell$ it is shown the distance of mean in units of standard deviation of the mean itself.

\begin{figure}[ht]
\centering
\begin{tabular}{c}
\includegraphics[width=103mm]{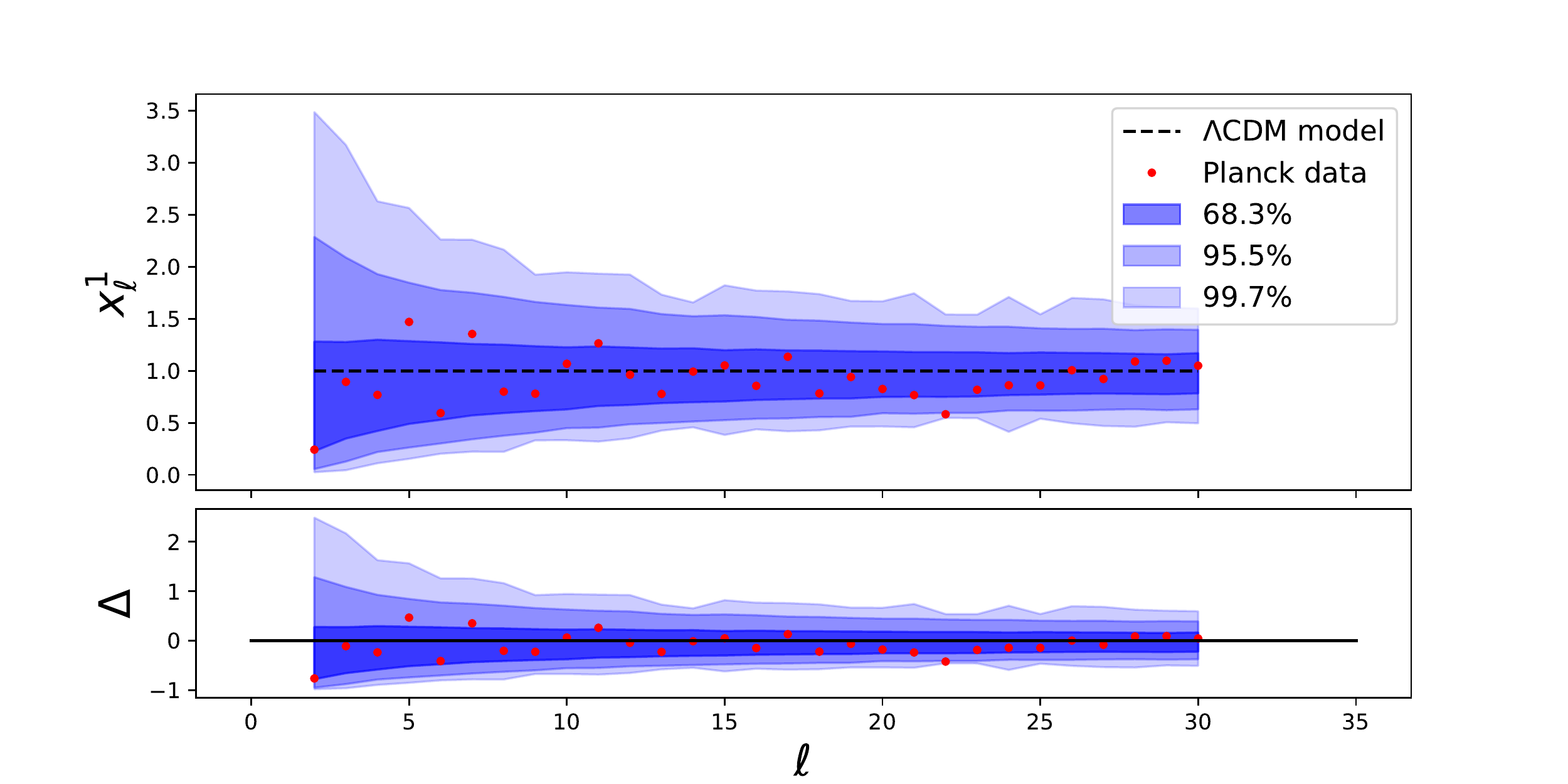}  \\
\includegraphics[width=103mm]{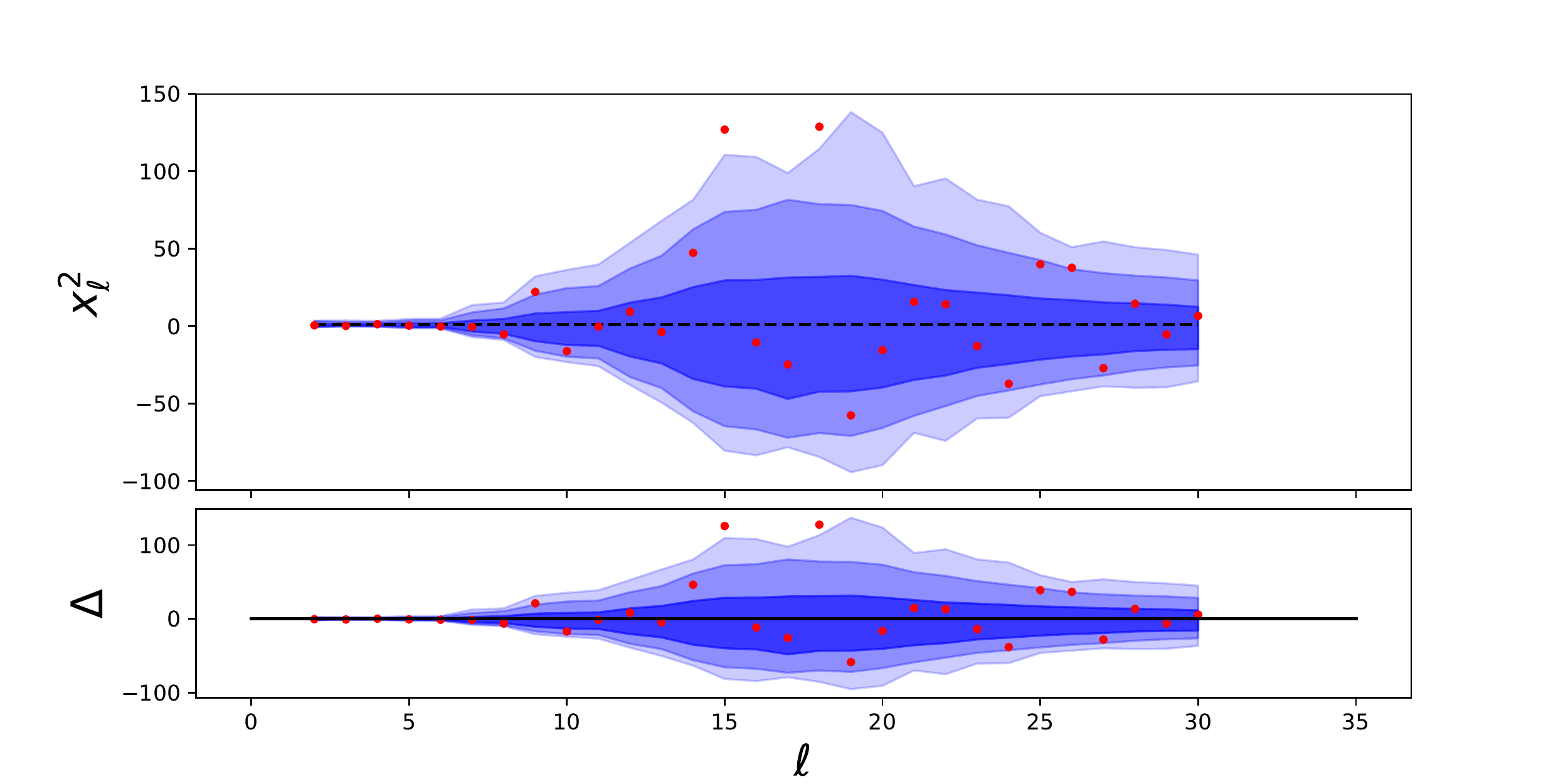} \\
\includegraphics[width=103mm]{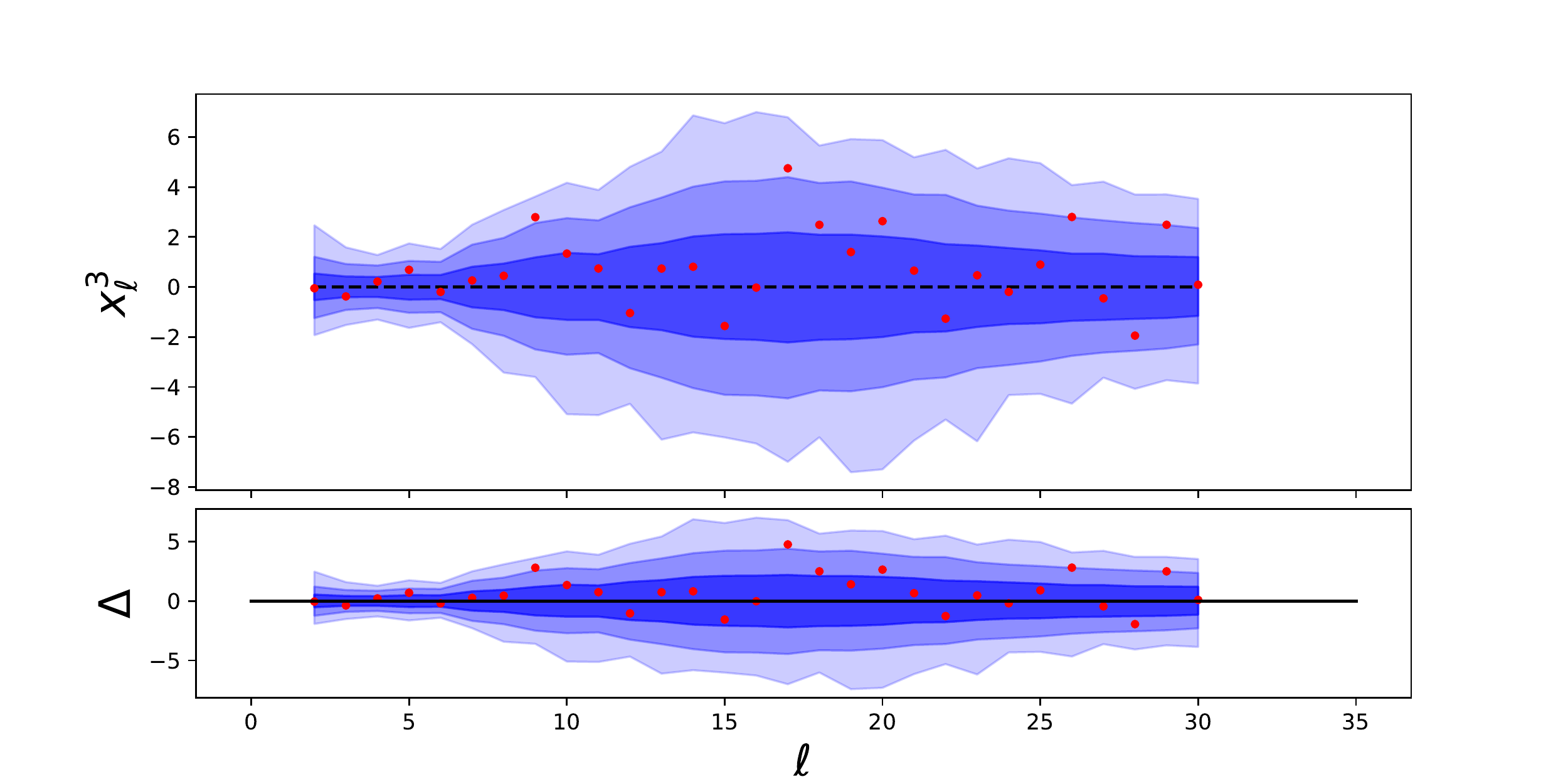}
\end{tabular}
\caption{\small Red dots represent $x^{(1)}_{\ell}$ (upper panel), $x^{(2)}_{\ell}$ (middle panel) and $x^{(3)}_{\ell}$ (lower panel) as a function of $\ell$ obtained from the {\sc Planck}-WMAP low-$\ell$ data set. Error bars (blue regions) represent the uncertainties associated to the estimates. Dashed horizontal lines represent what theoretically expected for 
the averages of $x^{(1)}_{\ell}$, $x^{(2)}_{\ell}$ and $x^{(3)}_{\ell}$, see equations (\ref{meanxi1}),(\ref{meanxi2}) and (\ref{meanxi1dotxi2}).
Each panel displays also a lower box where for each $\ell$ it is shown the distance of the estimates in units of standard deviation of the estimate itself.}
\label{fig:two}
\end{figure}

Figure \ref{fig:two} shows the low-$\ell$ estimates of $x^{(1)}_{\ell}$, $x^{(2)}_{\ell}$ and $x^{(3)}_{\ell}$ of the {\sc Planck}-WMAP low-$\ell$ data set (red dots), with the contours at one, two and three $\sigma$ as estimated from simulations (blue regions).

Simulations for a LiteBIRD-like noise level \cite{Matsumura:2013aja} are obtained following the same procedure as described above but dividing the polarisation part of the noise covariance matrix by a factor of $100$.

\begin{figure}[ht]
\centering
\begin{tabular}{c}
\includegraphics[width=110mm]{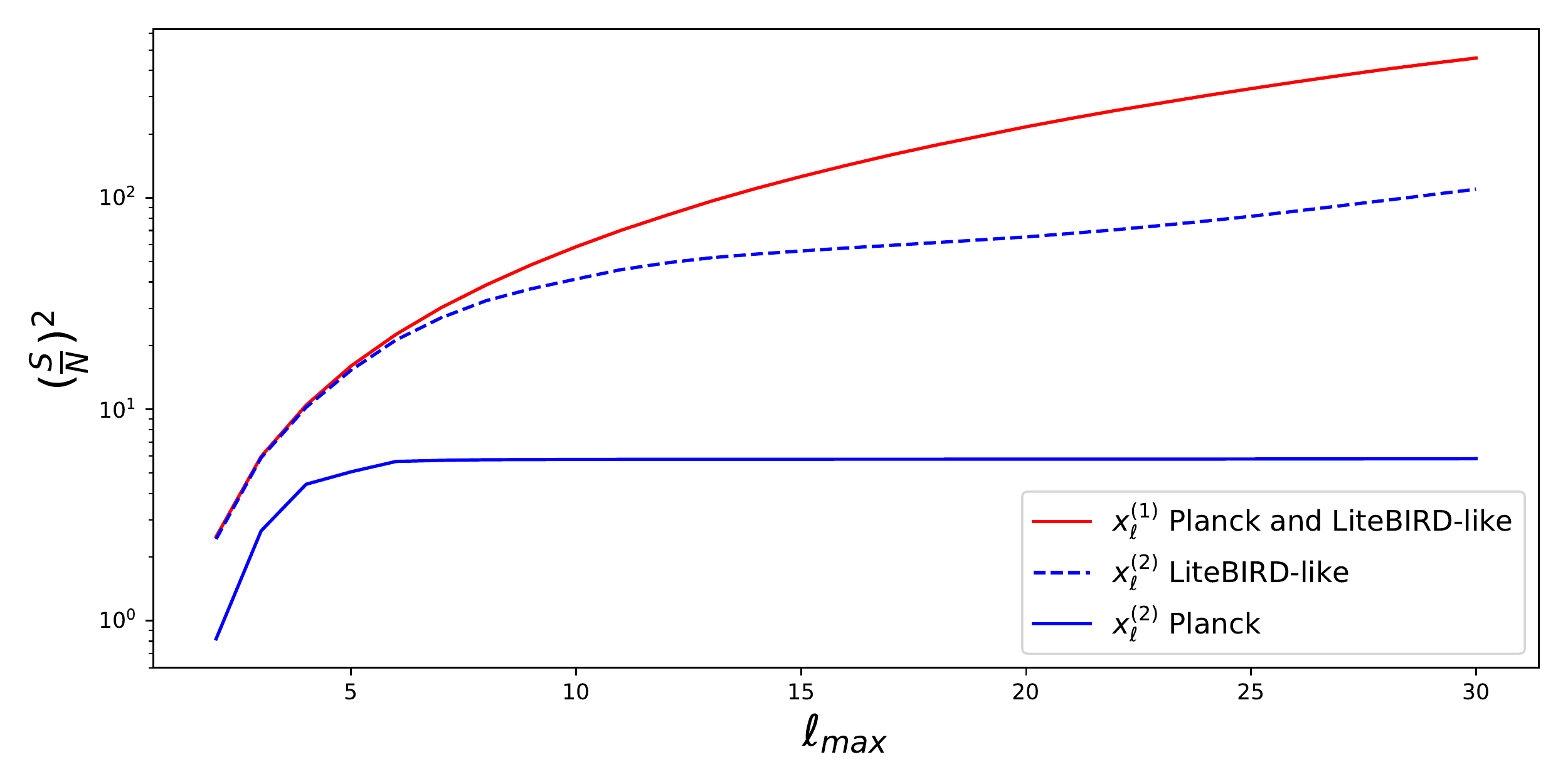}  \\
\end{tabular}
\caption{\small Signal-to-noise ratio of $x^{(1)}_{\ell}$ (red curve) and $x^{(2)}_{\ell}$ as a function of $\ell_{max}$ for the {\sc Planck}-WMAP low-$\ell$ data set (solid blue line) and for the LiteBIRD-like noise level (dashed blue line). 
While the signal contained in $x^{(1)}_{\ell}$ grows monotonically in the considered range, $x^{(2)}_{\ell}$ saturates at $\ell_{max} \sim 6$ for the {\sc Planck}-WMAP.
Instead the signal-to-noise ratio of $x^{(2)}_{\ell}$ for the LiteBIRD-like noise level grows monotonically,
and therefore, in this case, we can choose the maximum $\ell_{max}$ available in our simulations, i.e. $\ell_{max}=30$.}
\label{fig:three}
\end{figure}

The choice of the $\ell_{max}$ parameter, which enters the definition of $P$, see eq.(\ref{definitionofP}), is dictated by the signal-to-noise ratio of $x^{(2)}_{\ell}$ since $x^{(1)}_{\ell}$ is always signal dominated in the whole range considered. In Figure \ref{fig:three} we display the signal-to-noise ratio ($S/N$), see \ref{signal-to-noiseratio}, of the NAPS for the {\sc Planck}-WMAP low-$\ell$ data set (see solid lines). While for $x^{(1)}_{\ell}$ such a ratio grows monotonically (red line), for $x^{(2)}_{\ell}$ (solid blue line) it saturates around $\ell_{max} \sim 6$. Consequently we will employ the estimator $P$ with $\ell_{max}=6$ for the {\sc Planck}-WMAP low-$\ell$ data set.
%
The signal-to-noise ratio of $x^{(2)}_{\ell}$ for the LiteBIRD-like noise level is instead shown in Figure \ref{fig:three} as a dashed blue line. Since such a ratio grows monotonically, in this case we can choose the maximum $\ell_{max}$ available in our simulations, i.e. $\ell_{max}=30$.

Note that this limitation in the choice of  $\ell_{max}$ does not apply to $\tilde P$. 
In this case the coefficients $\alpha_{\ell}$ and $\beta_{\ell}$ adjust themselves automatically (depending on the signal-to-noise ratio) such that noise-dominated multipoles do not contribute to the estimator, see also Section \ref{optmised} and \ref{resultsPtilde}.

\section{Results of the analyses}
\label{results}

\subsection{Results for $P$}

\begin{figure}[ht]
\centering
\begin{tabular}{c}
\includegraphics[width=120mm]{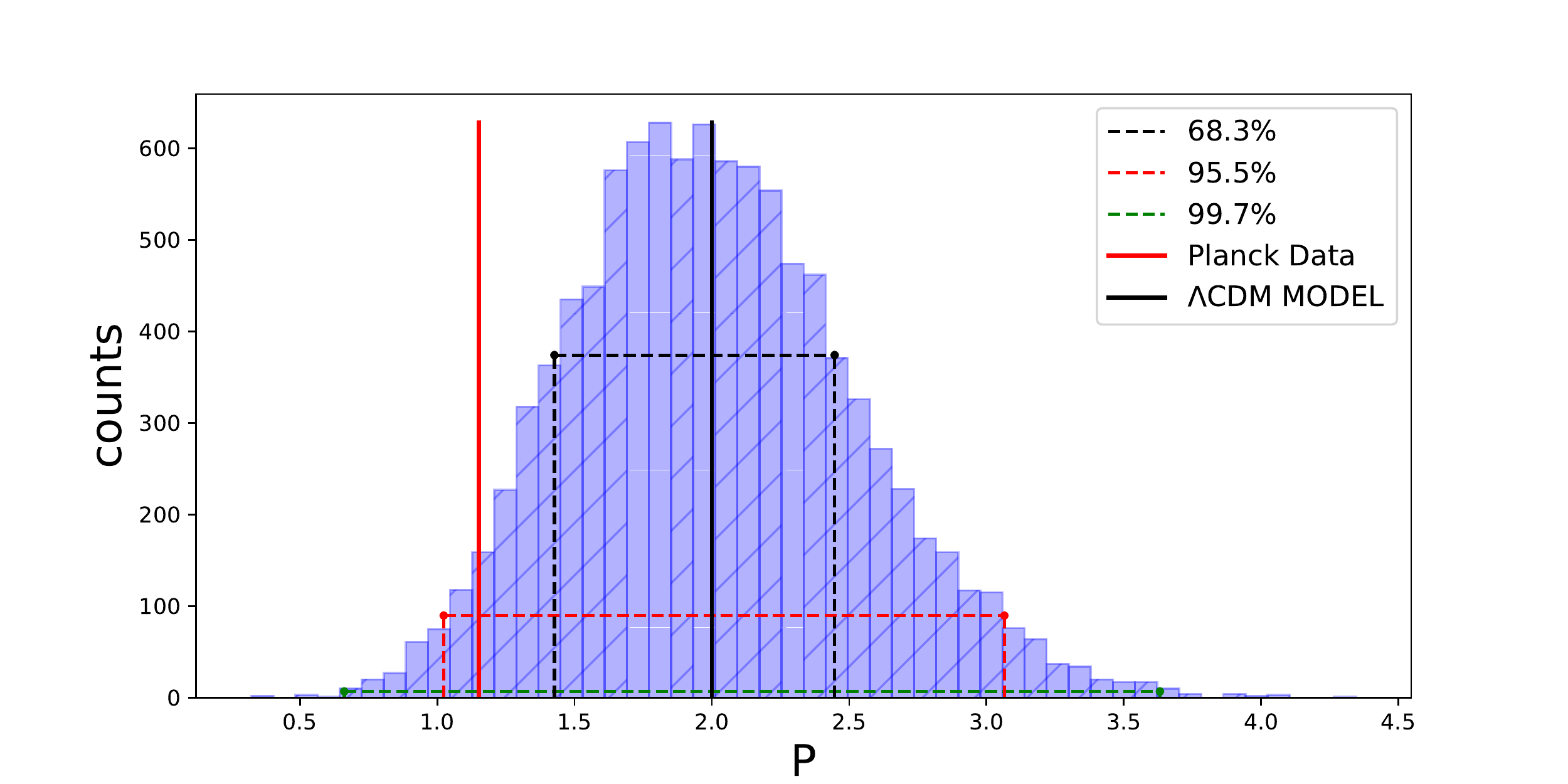}
\end{tabular}
\caption{\small Empirical distribution of the P estimator, see eq.~(\ref{definitionofP}), for $\ell_{max} = 6$. The red and the black vertical lines represent the values of the Planck data and of the $\Lambda$CDM model respectively. The black, red and green dashed lines indicate the boundaries of the 68.3\%, 95.5\% and 99.7\% confidence regions respectively.}
\label{fig:five}
\end{figure}

In Figure \ref{fig:five} we plot the empirical distribution expected in $\Lambda$CDM for $P$ with $\ell_{max}=6$ considering the {\sc Planck}-WMAP low-$\ell$ characteristics. The red vertical line stands for the observed value of the {\sc Planck}-WMAP low-$\ell$ data set. The lower-tail probability (LTP) of the observed value of $P$ is $3.63\%$. Such a value is smaller than the corresponding LTP of $P$ when, still with $\ell_{max}=6$, we neglect the contribution of $x^{(2)}_{\ell}$ in eq.~(\ref{definitionofP}). In that case the LTP we obtain is $17.63\%$. Similarly for the same maximum multipole, when we neglect the contribution of $x^{(1)}_{\ell}$ in $P$, we get a LTP of $6.71\%$.
In short, the combination of temperature and polarisation data provides a LTP smaller than what obtained with only temperature or only polarisation, although our findings cannot be considered as statistically anomalous. 

\begin{figure}[ht]
\centering
\begin{tabular}{c}
\includegraphics[width=110mm]{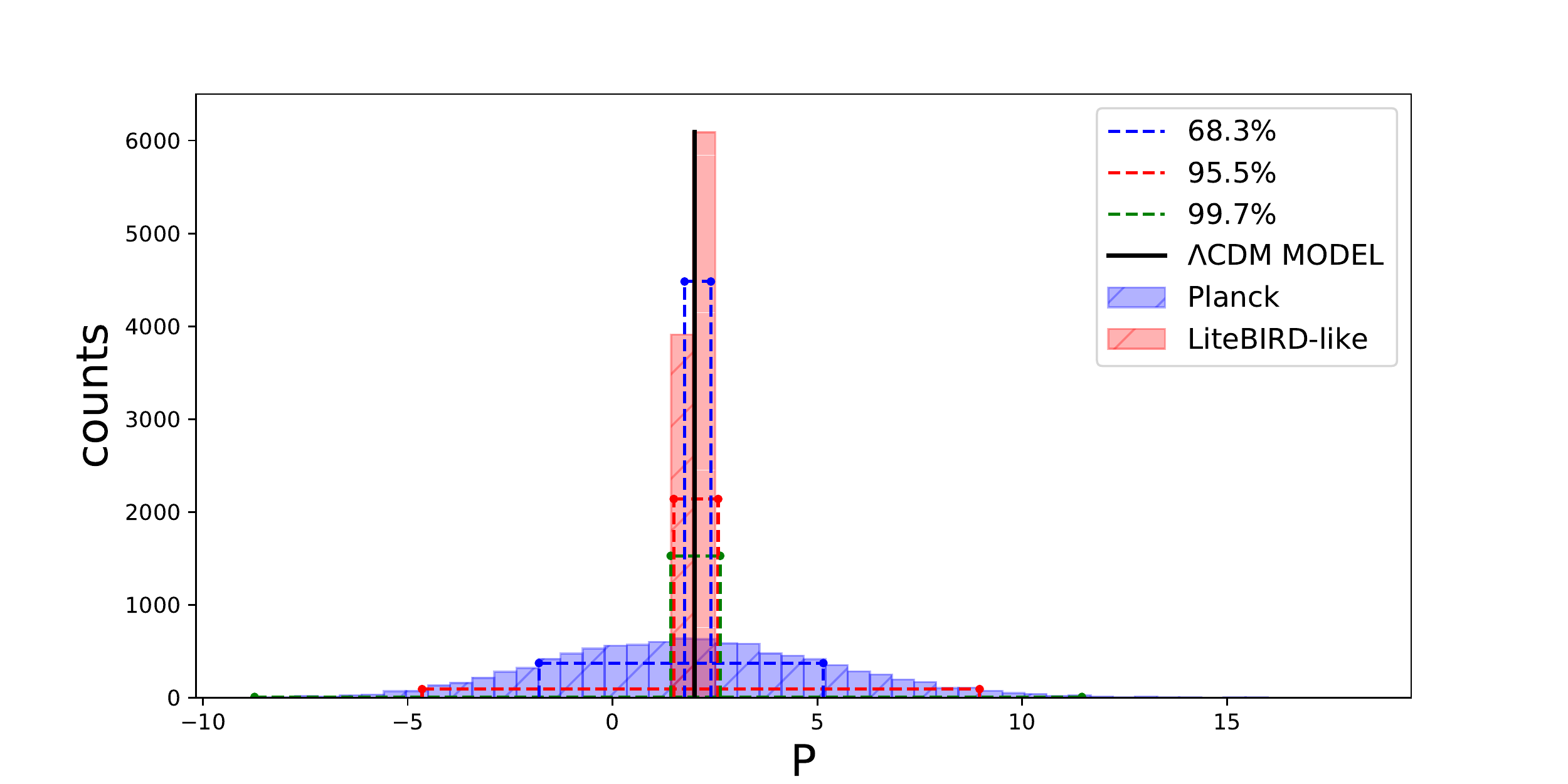} 
\end{tabular}
\caption{\small Empirical distribution of the P estimator, see eq.~(\ref{definitionofP}), for $\ell_{max} = 30$. Blue histogram refers to the {\sc Planck}-WMAP low-$\ell$ data set, while the red one is for the case of the LiteBIRD-like noise level.}
\label{fig:six}
\end{figure}

In Figure \ref{fig:six} we plot the empirical distribution of $P$ expected in $\Lambda$CDM with $\ell_{max}=30$ for the {\sc Planck}-WMAP low-$\ell$ data set and for the LiteBIRD-like noise level.
In order to evaluate the improvement of the latter with respect to the former, we build the ratios between the widths of the empirical distributions of $P$, corresponding to the level of $68.3\%$, for the LiteBIRD-like noise level ($\sigma_{LB}$) and for the {\sc Planck}-WMAP low-$\ell$ noise level ($\sigma_{Planck}$). 
We find that the width of the estimator $P$ in the LiteBIRD case can be even $30$ times smaller with respect to what obtained in the {\sc Planck}-WMAP low-$\ell$ case, if $\ell_{max} \gtrsim 20$.

\subsection{Results for $\tilde P$}
\label{resultsPtilde}

\begin{figure}[ht]
\centering
\begin{tabular}{c}
\includegraphics[width=110mm]{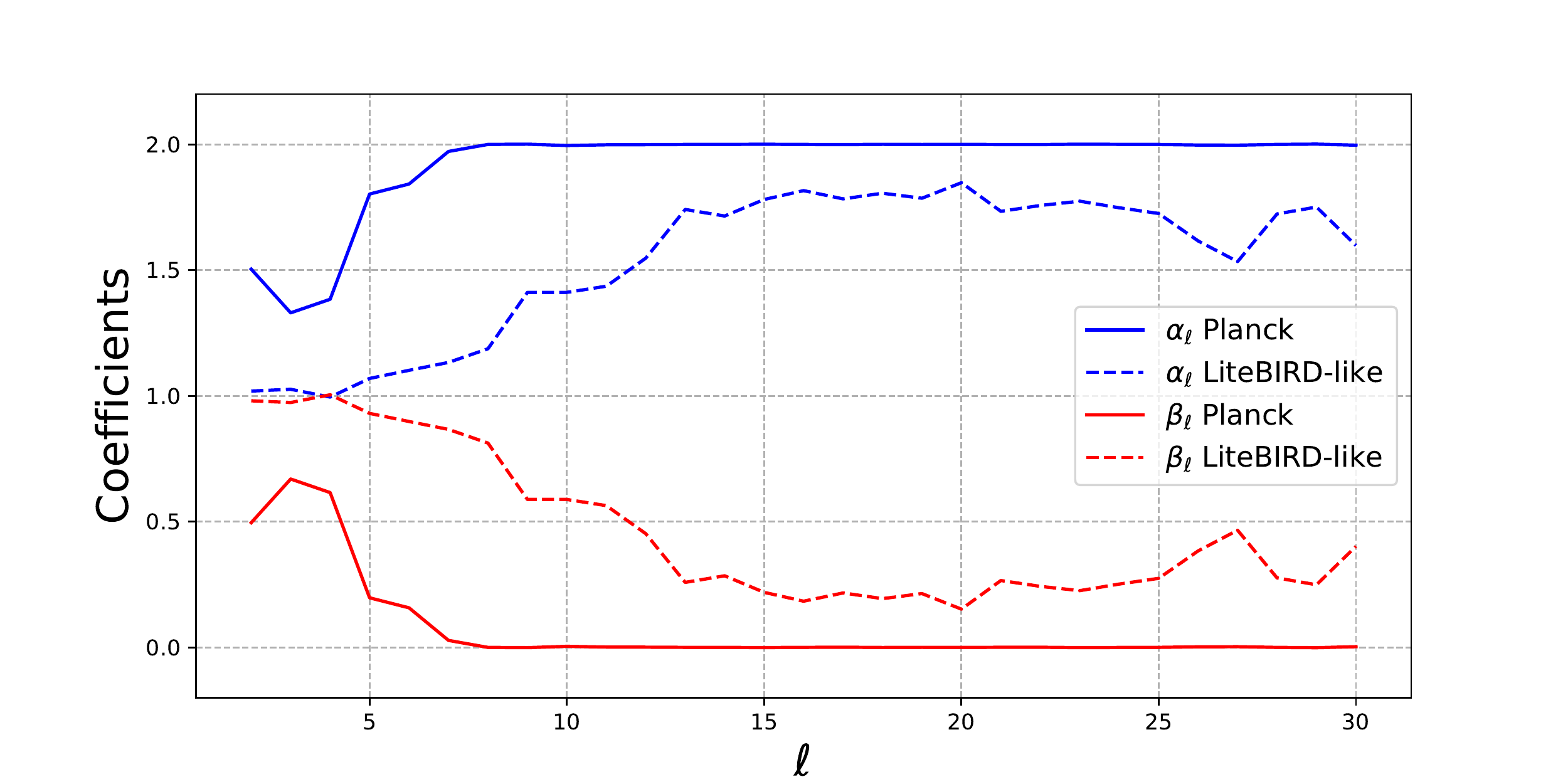} 
\end{tabular}
\caption{\small The behaviour of the coefficients $\alpha_{\ell}$ (blue line) and $\beta_{\ell}$ (red line), see eqs.~(\ref{definitionofPtilde}),(\ref{alphasol}) and (\ref{betasol}) for their definitions, as a function of $\ell$ for the {\sc Planck}-WMAP low-$\ell$ data set (solid lines) and for the LiteBIRD-like noise level (dashed lines).}
\label{fig:eight}
\end{figure}

\begin{figure}[ht]
\centering
\begin{tabular}{c}
\includegraphics[width=120mm]{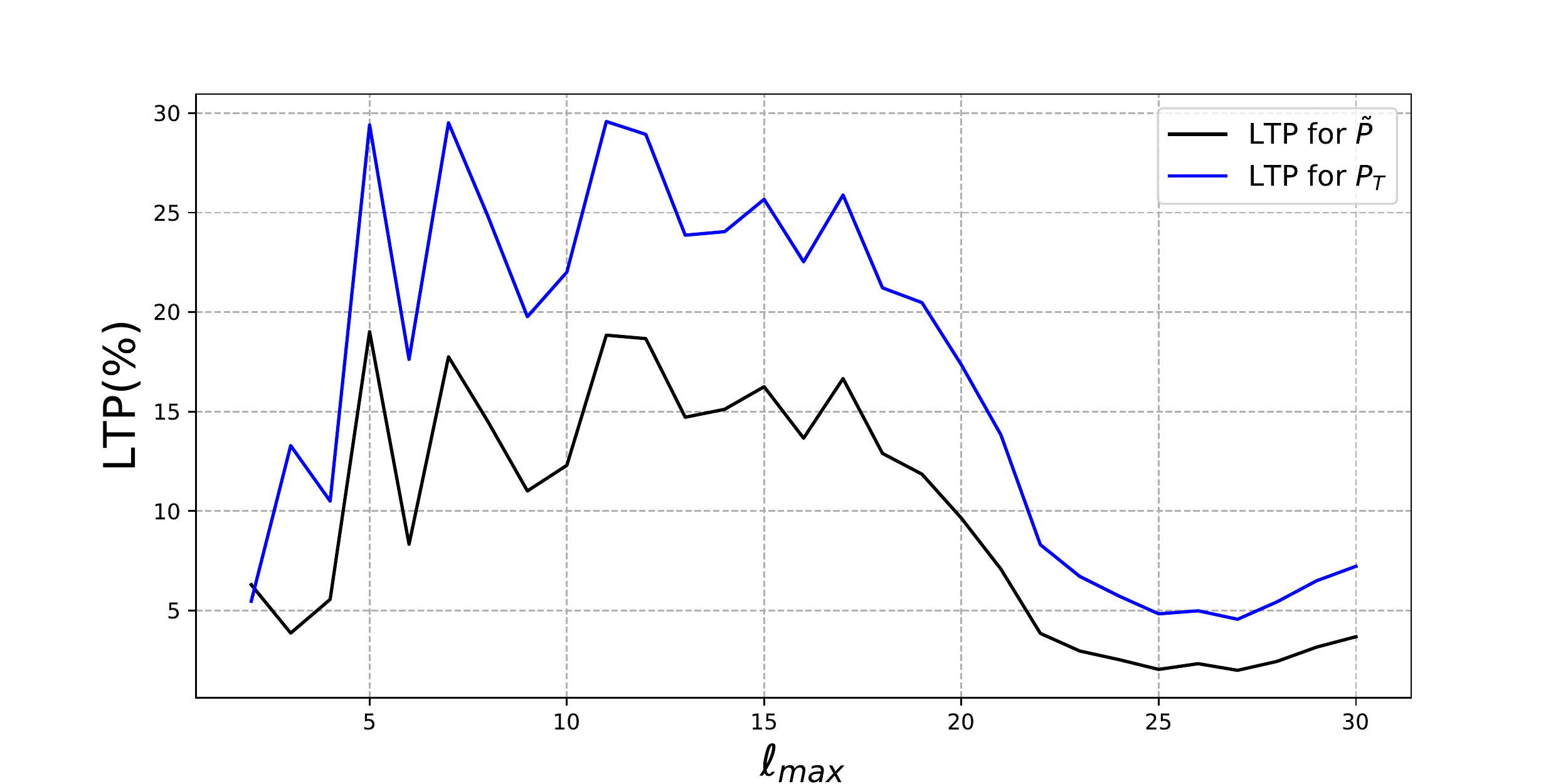} \\
\end{tabular}
\caption{\small LTP of $\tilde P$ (in black) and of $P_T$ (in blue) as a function of $\ell_{max}$ for the {\sc Planck}-WMAP low-$\ell$ data.}
\label{fig:nine}
\end{figure}

In Figure \ref{fig:eight} we plot $\alpha_{\ell}$ and $\beta_{\ell}$ (see equations (\ref{alphasol}) and (\ref{betasol})) as a function of $\ell$ for the {\sc Planck}-WMAP low-$\ell$ data set (solid lines).
Note how $\beta_{\ell}$ for $\ell > 7$ go to zero (and consequently $\alpha_{\ell} \rightarrow 2$ for same multipoles) because of the noise level in polarisation.
For $\ell_{max}=6$, even though the distribution of $\tilde P$ is $\sim 32\% $ narrower with respect to $P$ shown in Figure \ref{fig:five}, {\sc Planck}-WMAP low-$\ell$ data shift a little so that the LTP is increased to $8.33 \%$. 
However $\tilde P$, as already mentioned, is not limited in the choice of $\ell_{max}$ and still for the {\sc Planck}-WMAP low-$\ell$ data at $\ell_{max}=30$ we obtain a LTP at the level of $3.68 \%$.
In Figure \ref{fig:nine} we give the LTP for $\tilde P$ at each $\ell_{max}$, displayed in black, compared to a naive estimator defined only with temperature data as
$$ P_T = \frac{1}{(\ell_{max} -1)}\sum_{\ell=2}^{\ell_{max}}  x^{(1)}_{\ell} \, , $$ shown in blue.
It is interesting to note how the inclusion of the subdominant polarisation part impacts on the analysis making the LTP of $\tilde P$ smaller then $P_T$ for the whole $\ell-$range considered.
In particular for $P_T$ at $\ell_{max}=30$ we compute that the LTP is $7.22 \%$.

Still in Figure \ref{fig:eight} we plot $\alpha_{\ell}$ and $\beta_{\ell}$ as a function of $\ell$ for the LiteBIRD-like noise level (see dashed lines). 
Note that for this case none of the $\beta_{\ell}$ go to zero and therefore polarisation data provide a contribution for each of the multipoles
considered at large scale. Correspondently temperature data will not saturate the information entering $\tilde P$ for any considered multipoles.

In order to evaluate the impact of polarisation and temperature data on $\tilde P$ we define the following weights
\begin{eqnarray}
w_{x^{(1)}} (\ell_{max})= {1 \over 2 (\ell_{max}-1)} \sum_{\ell=2}^{\ell_{max}} \alpha_{\ell} \, , \\
w_{x^{(2)}} (\ell_{max})= {1 \over 2 (\ell_{max}-1)} \sum_{\ell=2}^{\ell_{max}} \beta_{\ell} \, ,
\end{eqnarray}
such that $w_{x^{(1)}} (\ell_{max}) + w_{x^{(2)}} (\ell_{max}) = 1$ for every $\ell_{max}$.
\begin{figure}[ht]
\centering
\begin{tabular}{c}
\includegraphics[width=120mm]{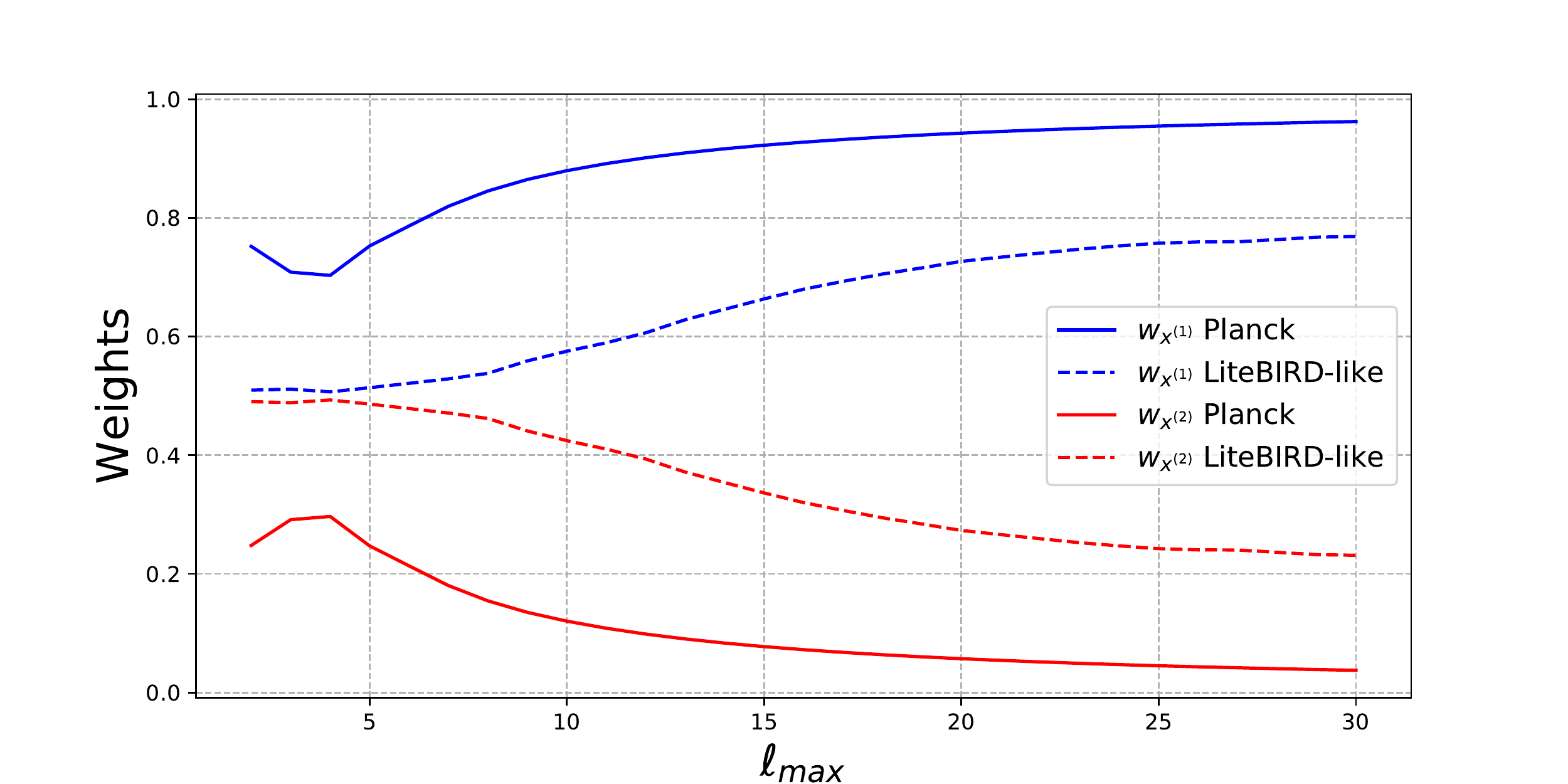} \\
\end{tabular}
\caption{\small $w_{x^{(1)}}$ (in blue) and $w_{x^{(2)}}$ (in red) as a function of $\ell_{max}$ for the {\sc Planck}-WMAP low-$\ell$ data set (solid) and the LiteBIRD-like case (dashed).}
\label{fig:ten}
\end{figure}
For $\ell_{max}=6$ we find that polarised {\sc Planck}-WMAP low-$\ell$ data contribute at the level of $21.4 \%$ to the building of $\tilde P$. This value increases to $47.9 \%$ for future LiteBIRD-like polarised data at the same maximum multipole. At $\ell_{max}=30$ we forecast that future LiteBIRD-like polarised data will weight as the $23.1\%$ with respect to the $3.8\%$ obtained with  {\sc Planck}-WMAP low-$\ell$ data, therefore providing an increasing factor $\sim 6$. The behaviour of $w_{x^{(1)}}$ and $w_{x^{(2)}}$ for each $\ell_{max}$ is given in Figure \ref{fig:ten}.

\begin{figure}[ht]
\centering
\begin{tabular}{c}
\includegraphics[width=120mm]{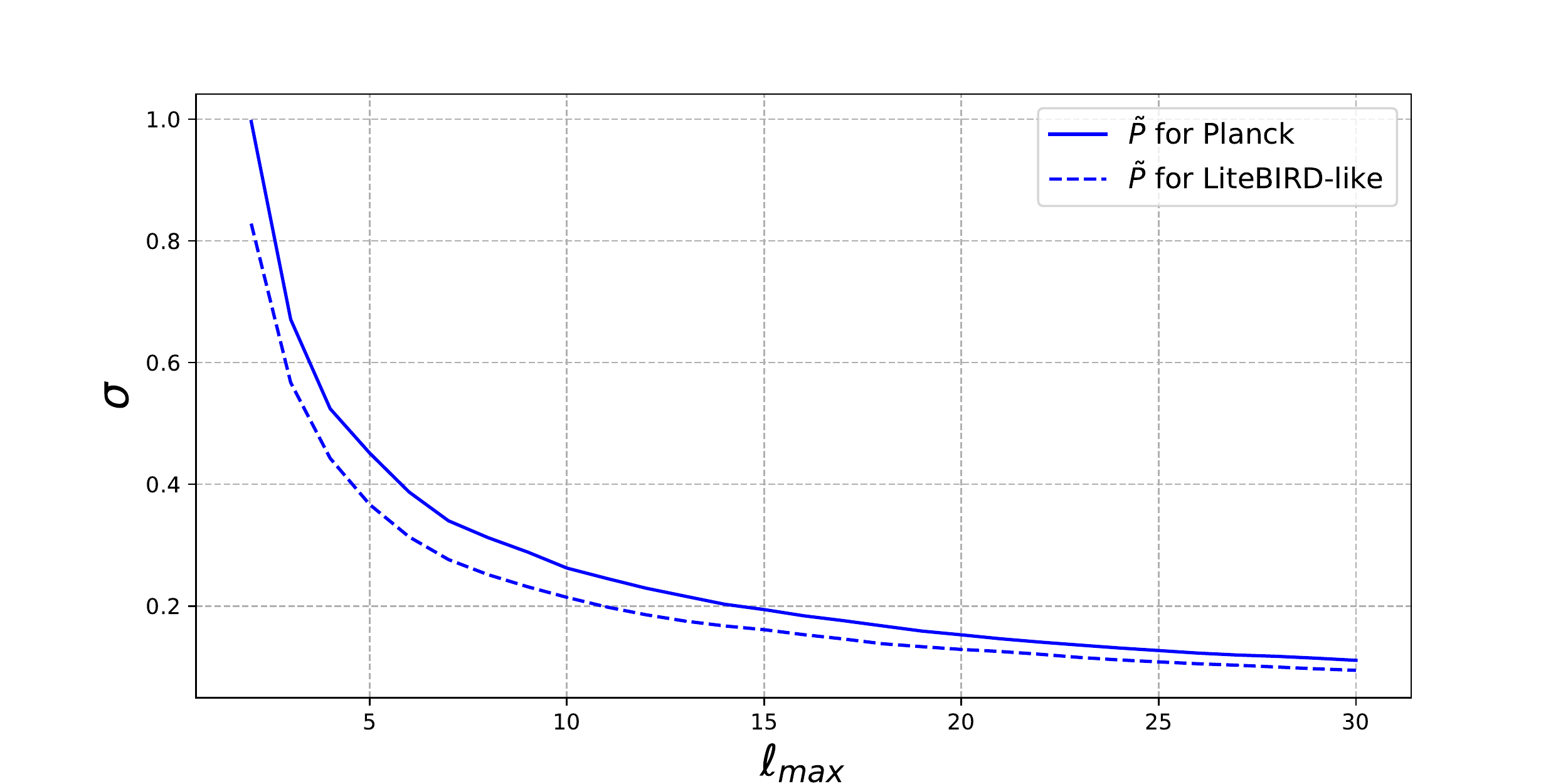} \\
\end{tabular}
\caption{\small Standard deviation $\sigma$ of $\tilde P$ versus $\ell_{max}$ for {\sc Planck}-WMAP low-$\ell$ data set (solid) and the LiteBIRD-like case (dashed).
}
\label{fig:eleven}
\end{figure}

We end this Section showing in Figure \ref{fig:eleven} how the standard deviation $\sigma$ of $\tilde P$ shrinks for each $\ell_{max}$ from current {\sc Planck} data (solid blue) to future LiteBIRD-like data (dashed blue). 
We compute that at low-$\ell$ future data will allow to build $\tilde P$ with a statistical uncertainty that will be around $20 \%$ smaller with respect to current {\sc Planck} data.

\section{Conclusions}
\label{conclusions}
In this paper we have proposed a new one-dimensional estimator, i.e. $P$ and its optimised version $\tilde P$, see equations (\ref{definitionofP}) and (\ref{definitionofPtilde}), which is able to jointly test the lack of power in TT, TE and EE. The main outcomes of this analysis are listed below.
\begin{enumerate}
\item Considering {\sc Planck}-WMAP low-$\ell$ data it is interesting to note that the inclusion of polarisation information through our new one-dimensional estimator, either $P$ or $\tilde P$, provides estimates which are less likely accepted in a $\Lambda$CDM model than the corresponding only-temperature version of the same estimator. In other words, polarisation though subdominant in terms of signal-to-noise ratio with respect to temperature, plays a non-negligible role in the evaluation of compatibility between data and the standard model. However the LTP obtained are at the level of few per cent and therefore still compatible with a statistical fluke. See for instance Figure \ref{fig:nine}: even though the weight of polarisation data is only around $4 \%$ of the total information budget, the LTP probability of $\tilde P$ is always smaller than its corresponding temperature-only version $P_T$. 
\item E-modes at large angular scale still contain information which might be capable to probe new physics beyond the standard cosmological model. 
	\begin{itemize}
		\item We forecast that future CMB polarised measurements \`a la LiteBIRD can tight the empirical distribution of $P$ up to a factor of $\sim30$.
		\item Considering the optimised version of the proposed estimator, i.e. $\tilde P$, we evaluate that future LiteBIRD-like measurements can shrink the statistical uncertainty by $23-17 \%$ and at the same time increasing the contribution of the polarisation part by a factor ranging from $\sim 2$ to $\sim 6$.
	\end{itemize}
\end{enumerate}

Future all-sky CMB experiments aimed at detecting primordial B-modes (which in turn are related to the energy scale of inflation) are designed to observe CMB polarisation with exquisite accuracy and precision. 
In order to make this possible, residual systematic effects both of instrumental and astrophysical origin 
have to be carefully measured or, at least, kept under control. In this paper we suppose that this is the case and that the statistical noise is the dominant source of uncertainty. 
Under these circumstances, E-modes will be in practice known at the cosmic variance limit at large angular scales. 
This is a great opportunity, since E-mode polarisation might contain important information 
about the lack of power anomaly currently observed only in the temperature map, which could be tracing new physical phenomena beyond the standard cosmological model in the early universe. 

\appendix

\section{Computation of the weights for $\tilde P$}
\label{optmisedweights}

We use the method of the Lagrange multipliers to minimise the variance of $\tilde P$, i.e. $var(\tilde{P})$, keeping fixed the 
expected value of $\tilde P$. This can be achieved requiring that 
\begin{equation}
\alpha_{\ell} + \beta_{\ell} = const = 2 \, ,
\label{constraint}
\end{equation}
for each multipole $\ell$.
Replacing the definition of $\tilde P$, see equation (\ref{definitionofPtilde}), in the expression of $var(\tilde{P})$, one obtains 
\begin{eqnarray}
var(\tilde{P})& \equiv &\langle (\tilde{P} - \langle \tilde{P} \rangle)^{2} \rangle = \langle (\tilde{P})^{2}\rangle - \langle \tilde{P} \rangle^{2} = \nonumber \\ 
&=& \sum_{\ell} var(\tilde{P_{\ell}}) 
\label{varP}
\end {eqnarray}
where the cross-terms among different multipoles goes exactly to zero in the full sky case, and with $var(\tilde{P_{\ell}})$ defined as
\begin{equation}
var(\tilde{P_{\ell}})=  \bar \alpha_{\ell} ^{2} var(x_{\ell}^{(1)}) + \bar \beta_{\ell}^{2} var(x_{\ell}^{(2)}) + 2 \bar \alpha_{\ell} \bar \beta_{\ell} cov(x_{\ell}^{(1)},x_{\ell}^{(2)}) \, ,
\label{varPell}
\end{equation}
where the barred quantities are defined as $\bar y = y / (\ell_{max}-1)$ and where $var(x_{\ell}^{(1)})$ and $var(x_{\ell}^{(2)})$ are the variance of $x_{\ell}^{(1)}$ and $x_{\ell}^{(2)}$ respectively, and $cov(x_{\ell}^{(1)},x_{\ell}^{(2)})$ is their covariance
\begin{equation}
cov(x_{\ell}^{(1)},x_{\ell}^{(2)}) = \langle(x_{\ell}^{(1)} - \langle x_{\ell}^{(1)} \rangle)(x_{\ell}^{(2)} -\langle x_{\ell}^{(2)}\rangle) \rangle.
\end{equation}
Because of Eq.~(\ref{varP}), the minimisation of $var(\tilde{P})$ is equivalent to the minimisation of each $var(\tilde{P_{\ell}})$.
As it is customary in the Lagrange multiplier method, for each multipole $\ell$ we introduce a new variable $\bar \lambda_{\ell}$, known as the Lagrange multiplier, and minimise the function $F(\bar \alpha_{\ell},\bar \beta_{\ell},\bar \lambda_{\ell})$
which is defined as
\begin{equation}
F(\bar \alpha_{\ell},\bar \beta_{\ell},\bar \lambda_{\ell}) = var(\tilde{P_{\ell}}) + \bar \lambda_{\ell} \left(\bar \alpha_{\ell} + \bar \beta_{\ell} -{2 \over (\ell_{max}-1)}\right) \, .
\end{equation}
This is equivalent to minimise the variance of $\tilde P_{\ell}$ on the constrain given by equation (\ref{constraint}) (multiplied by $1 / (\ell_{max}-1)$).
Therefore we compute the partial derivatives with respect to the coefficients $\bar \alpha_{\ell}$, $\bar \beta_{\ell}$ and $\bar \lambda_{\ell}$ and set them to be zero:
\begin{eqnarray}
\frac{\partial F(\bar \alpha_{\ell},\bar \beta_{\ell},\bar \lambda_{\ell})}{\partial \bar \alpha_{\ell}} & = & 2\bar \alpha_{\ell} \, var(x_{\ell}^{(1)}) + 2\bar \beta_{\ell} \, cov(x_{\ell}^{(1)},x_{\ell}^{(2)}) + \bar \lambda_{\ell} = 0, \\
\frac{\partial F(\bar \alpha_{\ell},\bar \beta_{\ell},\bar \lambda_{\ell})}{\partial \bar \beta_{\ell}} & = & 2\bar \beta_{\ell} \, var(x_{\ell}^{(2)}) + 2 \bar \alpha_{\ell} \, cov(x_{\ell}^{(1)},{x}_{\ell}^{(2)}) + \bar \lambda_{\ell} = 0, \\
\frac{\partial F(\bar \alpha_{\ell},\bar \beta_{\ell},\bar \lambda_{\ell})}{\partial \bar \lambda_{\ell}} & = & \bar \alpha_{\ell} + \bar \beta_{\ell} -{2 \over (\ell_{max}-1)} = 0 \, .
\end{eqnarray}
This set of equations is solved by equations (\ref{alphasol}), (\ref{betasol}) together with
\begin{eqnarray}
& & \bar \lambda_{\ell} = - {2 \over {(\ell_{max}-1)}} \left( \phantom{  {\left( var(x_{\ell}^{(3)}) \right)}  \over {var( x^{(4)}_{\ell} )} }   \!\!\!\!\!\!  \!\!\!\!\!\!  \!\!\!\!\!\!  \!\!\!\!\!\!  \!\!\!\!\!\! 
cov(x_{\ell}^{(1)},x_{\ell}^{(2)}) + \right. \nonumber \\ 
& & \left. \!\!\!\!\!\! + { var(x_{\ell}^{(1)}) var(x_{\ell}^{(2)}) - cov(x_{\ell}^{(1)},x_{\ell}^{(2)}) \left( var(x_{\ell}^{(1)}) + var(x_{\ell}^{(2)}) \right)  \over {var( x^{(1)}_{\ell} ) + var( x^{(2)}_{\ell} )-2cov(x^{(1)}_{\ell} ,x^{(2)}_{\ell})}} \right)  \, .
\end{eqnarray}

\section{Noise generation through Cholesky decomposition}
\label{choldecomp}

Given a noise covariance matrix $\mathbb{N}$ defined over the observed pixels, it is possible to generate a noise map, $m_n$, statistically compatible with $\mathbb{N}$, through the following expression
\begin{equation}
m_n = \mathbb{L} y \, ,
\end{equation}
where $\mathbb{L}$ is the lower triangular matrix of the Cholesky decomposition \cite{numerical} such that
\begin{equation}
\mathbb{N} = \mathbb{L} \mathbb{L}^t \, ,
\end{equation}
and $y$ is a vector with the same dimension as $m_n$ and whose entries are randomly extracted from a normal distribution.
In this way $m_n$ turns out to be statistically compatible with $\mathbb{N}$ since
\begin{equation}
\langle (m_n) (m_n)^t \rangle = \langle (\mathbb{L} y) (\mathbb{L} y)^t \rangle = \mathbb{L}  \langle y y^t \rangle \mathbb{L}^t =  \mathbb{L} \mathbb{I} \mathbb{L}^t = \mathbb{N} \, ,
\end{equation}
where $ \mathbb{I}$ is the identity matrix and $\langle ... \rangle$ stands for ensamble average.

\section{Signal-to-noise ratio}
\label{signal-to-noiseratio}
The total signal-to-noise ratio $\left(S/N\right)^{2}_{\ell_{max}}$ contained in $x^{(1)}_{\ell}$ or $x^{(2)}_{\ell}$ up to a maximum harmonic scale $\ell_{max}$, is defined summing up $\left(S/N\right)_{l}^{2}$ over the multipoles $\ell$ from $2$ to $\ell_{max}$ as:
\begin{equation}
\left(\frac{S}{N}\right)_{\ell_{max}}^{2} = \sum_{l=2}^{\ell_{max}} \left(\frac{S}{N}\right)_{l}^{2}
\label{s-n}
\end{equation}
where 
\begin{equation}
\left(\frac{S}{N}\right)_{l}^{2} = { \langle x^{(i)}_{\ell} \rangle^2 \over \langle \left( x^{(i)}_{\ell} - \langle x^{(i)}_{\ell} \rangle \right) ^2 \rangle} = { 1 \over \langle \left( x^{(i)}_{\ell} - 1 \right) ^2 \rangle} \, ,
\end{equation}
since $\langle x^{(i)}_{\ell} \rangle =1$, for $i=1,2$.

\paragraph{Acknowledgments}

We acknowledge the use of computing facilities at NERSC (USA), of the HEALPix package \cite{Gorski:2004by}, and of the {\sc Planck} Legacy Archive (PLA).
This research was supported by ASI through the Grant 2016-24-H.0 (COSMOS) and through the ASI/INAF Agreement I/072/09/0 for the Planck LFI Activity of Phase E2, and by INFN (I.S. FlaG, InDark). 
L.M. acknowledges the PRIN MIUR 2015 ``Cosmology and fundamental physics: illuminating the dark universe with Euclid''.


\section*{References}

\begin{thebibliography}{99}

\bibitem{Schwarz:2015cma}
  D.~J.~Schwarz, C.~J.~Copi, D.~Huterer and G.~D.~Starkman,
  Class.\ Quant.\ Grav.\  {\bf 33} (2016) no.18,  184001
  [arXiv:1510.07929 [astro-ph.CO]].

\bibitem{Muir:2018hjv}
  J.~Muir, S.~Adhikari and D.~Huterer,
  arXiv:1806.02354 [astro-ph.CO].

\bibitem{Monteserin:2007fv}
  C.~Monteserin, R.~B.~B.~Barreiro, P.~Vielva, E.~Martinez-Gonzalez, M.~P.~Hobson and A.~N.~Lasenby,
  Mon.\ Not.\ Roy.\ Astron.\ Soc.\  {\bf 387} (2008) 209
  [arXiv:0706.4289 [astro-ph]].

\bibitem{Cruz:2010ud}
  M.~Cruz, P.~Vielva, E.~Martinez-Gonzalez and R.~B.~Barreiro,
  Mon.\ Not.\ Roy.\ Astron.\ Soc.\  {\bf 412} (2011) 2383
  [arXiv:1005.1264 [astro-ph.CO]].

\bibitem{Gruppuso:2013xba}
  A.~Gruppuso, P.~Natoli, F.~Paci, F.~Finelli, D.~Molinari, A.~De Rosa and N.~Mandolesi,
  JCAP {\bf 1307} (2013) 047
  [arXiv:1304.5493 [astro-ph.CO]].

\bibitem{Ade:2013nlj}
  P.~A.~R.~Ade {\it et al.} [Planck Collaboration],
  Astron.\ Astrophys.\  {\bf 571} (2014) A23
  doi:10.1051/0004-6361/201321534
  [arXiv:1303.5083 [astro-ph.CO]].


\bibitem{Ade:2015hxq}
  P.~A.~R.~Ade {\it et al.} [Planck Collaboration],
  Astron.\ Astrophys.\  {\bf 594} (2016) A16
  [arXiv:1506.07135 [astro-ph.CO]].
  
\bibitem{Contaldi:2003zv}
  C.~R.~Contaldi, M.~Peloso, L.~Kofman and A.~D.~Linde,
  JCAP {\bf 0307} (2003) 002
  doi:10.1088/1475-7516/2003/07/002
  [astro-ph/0303636].
  
\bibitem{Cline:2003ve}
  J.~M.~Cline, P.~Crotty and J.~Lesgourgues,
  JCAP {\bf 0309} (2003) 010
  doi:10.1088/1475-7516/2003/09/010
  [astro-ph/0304558].
  
\bibitem{Destri:2009hn}
  C.~Destri, H.~J.~de Vega and N.~G.~Sanchez,
  Phys.\ Rev.\ D {\bf 81} (2010) 063520
  doi:10.1103/PhysRevD.81.063520
  [arXiv:0912.2994 [astro-ph.CO]].
  
  
\bibitem{Cicoli:2013oba}
  M.~Cicoli, S.~Downes and B.~Dutta,
  JCAP {\bf 1312} (2013) 007
  doi:10.1088/1475-7516/2013/12/007
  [arXiv:1309.3412 [hep-th]].
  
  \bibitem{augusto_cmbdepression}
  E.~Dudas, N.~Kitazawa, S.~P.~Patil and A.~Sagnotti,
  JCAP {\bf 1205} (2012) 012
  [arXiv:1202.6630 [hep-th]];
  N.~Kitazawa and A.~Sagnotti,
  JCAP {\bf 1404} (2014) 017
  [arXiv:1402.1418 [hep-th]],
  EPJ Web Conf.\  {\bf 95} (2015) 03031
  [arXiv:1411.6396 [hep-th]],
  Mod.\ Phys.\ Lett.\ {\bf A 30} (2015) no.28,  1550137
  [arXiv:1503.04483 [hep-th]].

  
\bibitem{Gruppuso:2015zia}
  A.~Gruppuso and A.~Sagnotti,
  Int.\ J.\ Mod.\ Phys.\ {\bf D 24} (2015) no.12,  1544008
  [arXiv:1506.08093 [astro-ph.CO]].

\bibitem{Gruppuso:2015xqa}
  A.~Gruppuso, N.~Kitazawa, N.~Mandolesi, P.~Natoli and A.~Sagnotti,
  Phys.\ Dark Univ.\  {\bf 11} (2016) 68
  [arXiv:1508.00411 [astro-ph.CO]].
  
\bibitem{Gruppuso:2017nap}
  A.~Gruppuso, N.~Kitazawa, M.~Lattanzi, N.~Mandolesi, P.~Natoli and A.~Sagnotti,
  Phys.\ Dark Univ.\  {\bf 20} (2018) 49
  doi:10.1016/j.dark.2018.03.002
  [arXiv:1712.03288 [astro-ph.CO]].
  
\bibitem{Suzuki:2018cuy}
  A.~Suzuki {\it et al.},
  arXiv:1801.06987 [astro-ph.IM].
  
\bibitem{Copi:2013zja}
  C.~J.~Copi, D.~Huterer, D.~J.~Schwarz and G.~D.~Starkman,
  Mon.\ Not.\ Roy.\ Astron.\ Soc.\  {\bf 434} (2013) 3590
  doi:10.1093/mnras/stt1287
  [arXiv:1303.4786 [astro-ph.CO]].
  
  
\bibitem{Aghanim:2015xee}
  N.~Aghanim {\it et al.} [Planck Collaboration],
  Astron.\ Astrophys.\  {\bf 594} (2016) A11
  doi:10.1051/0004-6361/201526926
  [arXiv:1507.02704 [astro-ph.CO]].

\bibitem{Lattanzi:2016dzq}
  M.~Lattanzi {\it et al.},
  JCAP {\bf 1702} (2017) no.02,  041
  doi:10.1088/1475-7516/2017/02/041, 10.1088/1475-
  [arXiv:1611.01123 [astro-ph.CO]].
  
\bibitem{Gorski:2004by}
  K.~M.~Gorski, E.~Hivon, A.~J.~Banday, B.~D.~Wandelt, F.~K.~Hansen, M.~Reinecke and M.~Bartelman,
  Astrophys.\ J.\  {\bf 622} (2005) 759
  doi:10.1086/427976
  [astro-ph/0409513].  http://sourceforge.net/projects/healpix/.
  
\bibitem{Matsumura:2013aja}
  T.~Matsumura {\it et al.},
  J.\ Low.\ Temp.\ Phys.\  {\bf 176} (2014) 733
  doi:10.1007/s10909-013-0996-1
  [arXiv:1311.2847 [astro-ph.IM]].
  
\bibitem{Gruppuso:2009ab}
  A.~Gruppuso, A.~De Rosa, P.~Cabella, F.~Paci, F.~Finelli, P.~Natoli, G.~de Gasperis and N.~Mandolesi,
  Mon.\ Not.\ Roy.\ Astron.\ Soc.\  {\bf 400} (2009) 463
  doi:10.1111/j.1365-2966.2009.15469.x
  [arXiv:0904.0789 [astro-ph.CO]].
  
 \bibitem{numerical}
 W.~H.~Press, S.~A.~Teukolsky, W.~T.~Vetterling, B.~P.~Flannery
 ``Numerical Recipes 3rd Edition: The Art of Scientific Computing'', 2007
 isbn: 0521880688, 9780521880688, Cambridge University Press, New York, NY, USA.
 
 


\end{thebibliography}
\end{document}